\documentclass[%
 reprint,
superscriptaddress,
 amsmath,amssymb,
 aps,
]{revtex4-2}

\usepackage{placeins}
\usepackage{float}
\usepackage{caption}
\usepackage{stfloats}
\usepackage{color}
\usepackage{soul}
\newcommand{\diff}{\text{d}}
\usepackage[compat=1.1.0]{tikz-feynman}
\usepackage{csquotes}
\usepackage{tikz}

\usepackage{graphicx}
\usepackage{dcolumn}
\usepackage{bm}

\usepackage[colorlinks,
            linkcolor=blue,
            citecolor=blue,
            urlcolor=blue]{hyperref}

\begin{document}

\preprint{APS/123-QED}

\title{Are Black Hole and Cosmological Horizon Thermodynamics Necessarily Inequivalent?} 

\author{Zachary J. Hoelscher}
\affiliation{Department of Physics and Astronomy, Vanderbilt University, Nashville, TN 37235, USA}
\email{zachary.j.hoelscher@vanderbilt.edu} 

\date{\today}

\begin{abstract}
It has recently been argued that cosmological horizon thermodynamics and black hole thermodynamics are generically inequivalent in any consistent, UV complete theory of quantum gravity, where it was suggested that this could be falsified by the identification of a counterexample. This conjecture was motivated in part by noting that cosmological horizons do not incorporate many of the convenient features found in the black holes described by string theory. In this work, we consider as a toy model a universe with a single non-rotating black hole and a positive cosmological constant. We work in the Nariai limit where the mass of the black hole is taken to its maximum possible value, and where we also assert that the quantum field theory living on this background is strictly parity invariant. In the semi-classical regime where we neglect instability of the exact Nariai equilibrium under quantum fluctuations, we find that the observer cannot readily distinguish the physics in the neighborhood of the black hole horizon from the physics in the neighborhood of the cosmological horizon. This does not prove that the two are generically equivalent in any universe, though it does suggest that they cannot be entirely unrelated in general. We propose that any consistent theory of quantum gravity that aims to describe black holes in a universe with a positive cosmological constant should provide a related microscopic origin for the leading semiclassical cosmological entropy and black hole entropy in the neighborhood of the particular limit considered here.  
\end{abstract}

\maketitle


\textit{Introduction ---} Quantum gravity in de Sitter is an outstanding mystery. While progress has been made in providing a microscopic account for the entropy of some idealized black holes in string theory~\cite{Strominger_1996}, the microscopic origin of cosmological entropy is little understood. There has been tremendous success in building a holographic correspondence in Anti de Sitter spacetime~\cite{Maldacena_1999}, with much recent progress in holography in flat spacetime via the Celestial~\cite{Pasterski_2021} and Carrollian holography programs, where the latter is now believed to arise as a flat-space limit of AdS/CFT~\cite{Romain_2026}. Following the success of the AdS/CFT correspondence, Strominger proposed a similar dS/CFT correspondence~\cite{Strominger_2001}, though this has not met with similar success. Part of the challenge is that de Sitter lacks a known stable realization in string theory~\cite{Obied_2018}, which would provide a UV-complete embedding for the framework. 

The many challenges associated with quantum gravity in de Sitter spacetime led Trivedi to conjecture that black hole thermodynamics and cosmological horizon thermodynamics are generically inequivalent~\cite{Trivedi_2026, Trivedi_2026b}. He argues that Gibbons and Hawking's work on de Sitter thermodynamics may not be well founded~\cite{Hawking_1977}, with a similar argument for the  derivation of the Friedmann equations from the Clausius relation, $\delta Q = T \diff S$~\cite{Cai_2005}, as well as the entropic gravity proposal of Verlinde~\cite{Verlinde_2011}. 

We refer the reader to~\cite{Banihashemi_2022} for an attempt to give cosmological thermodynamics a more rigorous footing, in a manner analogous to black hole thermodynamics, as well as a discussion of  Schwarzschild - de Sitter in~\cite{Bhattacharya_2016}. Susskind has also studied Schwarzschild - de Sitter, in particular the Nariai limit. He argues that the UV completion should be a matrix theory, similar to BFSS~\cite{Susskind_2023}. 

In this work we consider the case of a single non-rotating black hole in a universe with a positive cosmological constant. In the maximum-mass limit, we find that the observer cannot distinguish the physics in the neighborhood of the cosmological horizon from the physics in the neighborhood of the black hole horizon, where we demand that the quantum field theory living on the background is strictly parity-invariant. This case does not describe the real world, though it functions as a useful toy model for a thought experiment. This illustrates that while there are important differences between black hole and cosmological horizons, it is not clear that black hole and cosmological horizon thermodynamics are entirely disconnected in every case. This is conceptually interesting as part of an effort to understand what a consistent theory of quantum gravity in a universe with a positive cosmological constant would look like. 

\noindent \textbf{Our Core Claim:} While the Nariai limit does not obviously prove that the microscopic origins of the black hole and cosmological entropy are truly identical in this limit, it does require that they both yield indistinguishable semi-classical near-horizon physics (where we consider a universe in which the quantum field theory is strictly parity-invariant), in which case the semiclassical black hole and cosmological horizon thermodynamics are indistinguishable to the observer, so they cannot be entirely unrelated. Any purportedly consistent theory of quantum gravity which claims to generically separate the two (e.g.~in the context of the Thermodynamic Split Conjecture~\cite{Trivedi_2026, Trivedi_2026b}) must confront this. This point is our key novelty and core claim. 

\begin{figure}[!htbp]
\includegraphics[width=\textwidth, height=3.5 in]{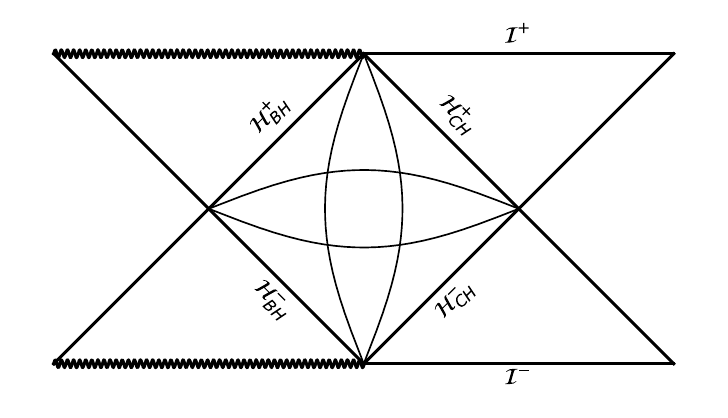}
\captionsetup{width=\textwidth, justification=raggedright} 
\caption{We show the Penrose diagram for the Schwarzschild - de Sitter spacetime, similar to the figure found in~\cite{Shi_2026}. Here the black hole mass is less than the extremal Nariai limit.} 
\label{fig:Penrose}
\end{figure}

As a thought experiment, we shall consider as a toy model a single non-rotating black hole in a universe with a positive cosmological constant. The Schwarzschild-de Sitter spacetime can be described by the following metric, where $G = c = 1$~\cite{Fernández-Silvestre_2022}, also see Figure~\ref{fig:Penrose}.

\vspace{-0.2 cm}

\begin{align}
  \diff s^2 &= -f(r) \diff t^2 + \frac{1}{f(r)} \diff r^2 + r^2 \diff \Omega^2\\
  f(r) &= 1 - \frac{2M}{r} - \frac{\Lambda r^2}{3}
\end{align}

\vspace{-0.2 cm}

One can now imagine scaling the mass of the black hole until the horizons coincide, then zooming into the region between the horizons. (It is important that the limits are taken in this order, in order to arrive at this particular metric~\cite{Guerrero_2022}.) This is known as the \enquote{Nariai limit}, which reduces to $\diff S_2 \times S_2$, given in the equation below~\cite{Nariai_1950, Nariai_1951, Ginsparg_1983, Stotyn_2015}. One may object that this is not a bona fide black hole, since it lacks a singularity, though we point out that this is the spacetime seen by the observer in the static patch in the limit that the black hole mass is taken to its maximal value~\cite{Shi_2026}. Since this describes the static patch, it is still the relevant metric for our problem. As it is indirectly relevant to our problem here, we note that an analogue of the proposed Kerr/CFT correspondence~\cite{Guica_2009} has been considered for a rotating analogue of the Nariai limit, though the authors note that the validity of the Cardy formula in this particular context is unclear, so the conclusions therein are not rigorous~\cite{Anninos_2010}.

\vspace{-0.2 cm}

\begin{equation}
  \diff s^2 = -\biggl(1 - \frac{\rho^2}{r_0^2}\biggr) \diff t^2 + \frac{\diff \rho^2}{1 - \frac{\rho^2}{r_0^2}} + r_0^2 \diff \Omega^2
\end{equation} 

\vspace{-0.2 cm}

In this limit, the black hole and cosmological horizons are located at $\rho=-r_0$ and $\rho=r_0$. The two horizons have equal area, equal entropy, and equal Hawking temperature. One can demonstrate that the Hawking temperatures are equal (without reference to the surface gravity) by appealing to the tunneling method~\cite{Belgiorno_2017}, which is a relatively general framework that is also known to work in FRW spacetimes for computing the Hawking temperature~\cite{Cai_2009}. We also refer the reader to~\cite{Eune_2013} and~\cite{JinHoCho_2008} for further discussions of entropy and thermodynamics in the Nariai spacetime. 

One can exchange the neighborhood of one horizon for the neighborhood of the other horizon by taking $\rho \to -\rho$~\cite{Stotyn_2015, Flam_2025}. The metric is also invariant under this transformation, as well as general parity transformations.

This discrete symmetry suggests that the physics of a strictly parity-invariant process near the black hole horizon can be directly related to a strictly parity-invariant process in the neighborhood of the cosmological horizon, and vice versa. For the purpose of this argument, which requires only the identification of a single example in which black hole and cosmological horizon thermodynamics are equivalent, we thus consider as a simplifying case a universe which does not incorporate weak interactions, which could suggest a gauge group such as $SU(3) \times U(1)$ with $\theta = 0$ in $\mathcal{L}_\theta = \theta \frac{g^2}{32\pi^2} G_{\mu\nu}^a \widetilde{G}^{a\mu\nu}$, rather than $SU(3) \times SU(2) \times U(1)$. We do not see an obvious reason why a UV complete theory of quantum gravity could not plausibly admit such a construction. (We note that existing work studies a universe which lacks weak interactions; interestingly, they find that it can support realistic behaviors, such as Big Bang nucleosynthesis, star formation, and structure formation. They suggest that this could arise as an EFT if the Higgs mass is of $\mathcal{O}(M_{PL})$~\cite{Harnik_2006}.) This suggests that black hole and cosmological horizon thermodynamics are not totally disconnected as one approaches the Nariai limit, which should be respected by any consistent theory of quantum gravity that claims to describe both black holes and a universe with a positive cosmological constant. 

We note that the identification of this spacetime as a finite-volume region between two horizons has been disputed, as Stotyn claims the horizons are more accurately regarded as degenerate in the exact coincidence limit. He motivates this from the observation that the isometry group of Nariai is enhanced relative to that of Schwarzschild-de Sitter~\cite{Stotyn_2015}. While this may at first appear to undermine the central thought experiment considered here, we point out that the traditional interpretation of Nariai as the limiting case of Schwarzschild - de Sitter (with a finite volume separating the horizons) has since been given stronger support through work which primarily seeks to bolster the argument that the Euclidean Nariai solution can be used to compute the black hole nucleation rate~\cite{Shi_2026}. 

A universe with a positive cosmological constant can undergo black hole \enquote{nucleation}, where a black hole is spontaneously formed as the result of a rare Boltzmann fluctuation. The Euclidean Nariai solution is an instanton of the gravitational path integral, and can be used to compute the following nucleation rate~\cite{Shi_2026}. Nariai naturally describes nucleation because it is the only smooth compact saddle point of the Schwarzschild - de Sitter gravitational path integral~\cite{Shi_2026}. 

\vspace{-0.4 cm}

\begin{equation}
  \Gamma_{\mathrm{NARIAI}} \sim \exp \biggl[-({I_{\mathrm{NARIAI}} - I_{\diff S})\biggr]} \sim \exp \biggl(-\frac{\pi}{G \Lambda} \biggl)
\end{equation}

\vspace{-0.1 cm}

While the probability of such an event is never zero, it is astoundingly small at late times, given small $\Lambda$. At early times, however, the situation is different, as nucleation can be relevant for the formation of primordial black holes during inflation~\cite{Bousso_1995, Shi_2026}. 

One of the key contributions of~\cite{Shi_2026} is to point out that one must include an observer in the Nariai spacetime, which they represent using a geodesic of a massive particle. This observer sits at the center of the static patch, on a geodesic where the cosmological expansion is balanced by the gravitational attraction from the black hole. For this observer, the enhanced symmetry $SO(2,1) \times SO(D-1)$ is reduced to $SO(1,1) \times SO(D-1)$, which is the isometry group of Schwarzschild - de Sitter~\cite{Shi_2026}. The observer which is necessary to enable a consistent interpretation of Nariai as encoding the nucleation rate is the exact observer which we care about for our thought experiment in this work. This observer thus sees Schwarzschild - de Sitter become $\diff S_2 \times S_2$ in the extremal-mass limit, where the black hole and cosmological horizon are separated by a finite proper distance, and where the horizons have equal Hawking temperatures at the exact Nariai equilibrium~\cite{Shi_2026}. Susskind also argues that the horizons remain separated by a finite proper distance in the Nariai limit, despite the fact that the radii coincide in the original coordinates~\cite{Susskind_2023}. 

We also point out that the traditional interpretation that the spacetime between the horizons can be blown up to yield $\diff S_2 \times S_2$ is simpler than the point of view suggested by Stotyn~\cite{Stotyn_2015} in the sense that the Schwarzschild-de Sitter entropy is continuous through the Nariai limit. If we regard the entropy as $S = (A_{BH} + A_C)/4G$, we can obtain a Nariai limit of $2A_{C}/4G$ from $\diff S_2 \times S_2$, whereas if the horizons merge into one, we might naively expect $A_{C}/4G$, which should be explained. This particular question is beyond the scope of this work.

While it is noteworthy that the black hole and cosmological horizon have equal Hawking temperature in the Nariai limit, this is not the only interesting comparison one can make. Green's functions have been computed in Nariai~\cite{Aalsma_2026}, as have absorption coefficients. In the Nariai limit, the scalar field absorption coefficients for the black hole and cosmological horizons are identical, yielding the following quantity~\cite{Shi_2026}.
\begin{equation}
  \Gamma_{\omega} = \frac{2 \sinh^2(\pi \omega)}{\cos\biggl(\pi \sqrt{1-4 \mu_{\ell}^2}\biggr) + \cosh(2 \pi \omega)}
\end{equation}
\vspace{-0.5 cm}
\begin{equation}
 \mu_{\ell} = \frac{m^2}{\Lambda} + \ell(\ell + 1)
\end{equation}

We also mention that while the exact Nariai limit is not plagued by large quantum gravitational corrections, it is believed to be an unstable equilibrium once quantum fluctuations are considered~\cite{Ginsparg_1983, Susskind_2023}, as a perturbation which makes the black hole hotter than the cosmological horizon breaks the thermodynamic equilibrium and allows the black hole to begin to evaporate. This does not invalidate our arguments, however. We are not claiming that black hole and cosmological horizon thermodynamics are always equivalent for any case and on any arbitrary timescale; instead, we argue that they cannot be wholly disconnected in every case. In order to make that weaker claim, we need only find one point in the solution space where the thermodynamics are equivalent, which we have done. Stability of this equilibrium is not necessary for this connection to exist. 

One could also possibly consider a case where the black hole and cosmological horizons remain separated by $\epsilon > 0$ in the original coordinates, and thus formally have slightly different temperatures, with a slightly broken reflection symmetry, but where $k_B |T_{BH} - T_C| <<<  \Lambda_{IR}$, where $\Lambda_{IR}$ is the IR cutoff. This case is relevant for small deviations from the exact Nariai limit. Here we assume that the differences imposed by $\epsilon>0$ are sufficiently small that the observer located between the horizons cannot readily tell the horizons apart in their effective theory. If the observer watches the system for some finite time window, the observer cannot resolve arbitrarily low frequencies, which correspondingly imposes an infrared cutoff. Unlike the exact Nariai limit, this is merely an effective description; however, rather than an exact equivalence, though it is relevant when one perturbs slightly away from the exact equilibrium we consider here. Since we are interested in the exact Nariai limit, we do not consider this case in further detail here.

The Thermodynamic Split Conjecture has recently proposed that black hole and cosmological horizon thermodynamics are generically inequivalent, with the suggestion that this can be falsified through the identification of a counter-example~\cite{Trivedi_2026, Trivedi_2026b}. As a thought experiment, we consider the maximal-mass limit of a single Schwarzschild black hole in a universe with a positive cosmological constant, where the quantum field theory living on this background respects parity invariance. While the individual ingredients in this work have existed separately, this work is novel for combining them in the context of the Thermodynamic Split Conjecture. We find that the observer cannot differentiate between the two horizons, suggesting that the underlying physics cannot be wholly unrelated in this limit. While the microscopic origin of de Sitter entropy remains an outstanding mystery, this still serves to indirectly shed a bit of light on the problem. We expect that this limit provides a useful test case for any putative theory of quantum gravity which aims to describe a universe with a positive cosmological constant.  

\noindent \textbf{Acknowledgments:} The author thanks Oem Trivedi, Mirko Landivar Gareca, Robert Scherrer, and Thomas Kephart for helpful discussions. The author was financially supported by a doctoral stipend at Vanderbilt University. 

\bibliography{apssamp}

\end{document}